\documentclass{aa}

\usepackage{graphics}

\def \dshism{${(\rm{D/H})}_{ISM}$}
\def \hii{\ion{H}{ii}}
\def \hi{\ion{H}{i}}
\def \di{\ion{D}{i}}
\def \nii{[\ion{N}{ii}]}
\def \oi{\ion{O}{i}}

\def \ha{H$\alpha$}
\def \hb{H$\beta$}

\def \da{D$\alpha$}
\def \db{D$\beta$}

\def \kms{${\rm km}\,{\rm s}^{-1}$}

\def \eg{{\it e.g.}}

\def \cm2{cm$^2$}
\def \lybd{Ly$\beta_{{\rm D}}$}
\def \lyb{Ly$\beta_{{\rm H}}$}

\begin{document}

   \thesaurus{08             
             (02.12.1        
              02.12.2        
              09.01.2        
              09.08.1        
              09.09.1 M42    
              12.03.3)}      
              
%
%

   \title{Detection of deuterium Balmer lines in the Orion Nebula
\thanks{Based on observations collected at the Canada-France-Hawaii
Telescope, Hawaii, USA.}
}

  \author{      G.~H\'ebrard \inst{1}
          \and
                D.~P\'equignot \inst{2}
          \and
                A.~Vidal-Madjar \inst{1}
          \and
                J.~R.~Walsh \inst{3}
          \and
                R.~Ferlet \inst{1}
          }

   \offprints{Guillaume H\'ebrard}

   \institute{Institut d'Astrophysique de Paris, CNRS,
              98 bis Boulevard Arago, F-75014 Paris, France 
              (hebrard@iap.fr, vidalmadjar, ferlet).
         \and
              Laboratoire d'Astrophysique Extragalactique et de 
              Cosmologie associ\'e au CNRS (UMR 8631) et \`a l'Universit\'e 
              Paris 7, DAEC, Observatoire de Paris-Meudon, F-92195 
              Meudon C\'edex, France (daniel.pequignot@obspm.fr).
         \and
              Space Telescope European Co-ordinating Facility, 
              European Southern Observatory, Karl-Schwarzschild-Strasse 2,
              D-85748 Garching bei M\"unchen, Germany 
              (jwalsh@eso.org).
             }

   \date{Received ? / Accepted ?}

   \maketitle

   \begin{abstract}

The detection and first identification of the deuterium Balmer 
emission lines, \da\ and \db, in 
the core of the Orion Nebula is reported. 
These lines are very narrow, have identical 
11~\kms\ velocity shifts with respect to \ha\ and \hb, 
are probably excited by UV continuum 
fluorescence from the Lyman (\ion{D}{i}) lines and arise from the 
interface between the \hii\ region and the molecular cloud.

      \keywords{Line: formation --
                Line: identification --
                H\ts {\sc ii} regions --
                ISM: individual objects: M42 --     
                ISM: atoms --
                Cosmology: observations
               }
   \end{abstract}

%

\section{Introduction}

Deuterium is believed to be entirely produced in the Big Bang 
and then steadily destroyed by astration 
(Epstein et al.~\cite{epstein76}). 
Standard models predict a decrease of 
its abundance by a factor 2--3 in 15~Gyrs
(\eg,~Tosi et al.~\cite{tosi98}). This picture is essentially 
constrained by deuterium abundance determinations at $\sim15$~Gyrs 
(primordial intergalactic clouds), 4.5~Gyrs (protosolar) and 0.0~Gyrs 
(interstellar medium). 
Although the evolution of the deuterium abundance seems to be 
qualitatively understood, the measurements 
show some dispersion. 
Thus, absorption in the Lyman series provides 
interstellar deuterium abundance 
\dshism$\simeq1.5\times10^{-5}$ 
(Linsky~\cite{linsky98}), 
but with fluctuations that may well be real 
(Vidal-Madjar et al. 1998). 
These dispersions led to the development of non-standard 
models in which, for example, deuterium may either decrease 
by more than a factor 4 
in 15~Gyrs (\eg,~Vangioni-Flam et al.~\cite{flam94}) 
or be created/destroyed by new mechanisms
[\eg, Lemoine et al.~(\cite{lemoine99}) for a~review].

A detailed appraisal of the evolution of deuterium is 
crucial for cosmology and galactic chemical evolution. 
The most reliable estimate of \dshism\ to date is based on 
far-UV observation from space (Copernicus, IMAPS, HST or FUSE) 
of the Lyman lines of D and H in absorption. 
These lines are also observed in the optical and near-UV 
to obtain D/H in high redshift quasar 
absorbers. Other D/H determinations include {\it in situ} measurements 
in the Solar System (\eg,~Mahaffy et al.~\cite{mahaffy98}), 
observations of molecules such as HD 
or DCN (\eg,~Bertoldi et al.~\cite{bertoldi99}) 
and observations of \ion{D}{i}~92~cm 
(\eg,~Chengalur et al.~\cite{chengalur97}).

New methods to determine D/H are of interest. One possibility 
is ground-based observation of the deuterium lines. 
The isotope shift of the deuterium Balmer lines 
with respect to the hydrogen Balmer lines 
is $-81.6$~\kms. These \di\ lines have never been identified before. 
Attempts to detect \da\ 
in absorption in the Sun (Beckers~\cite{beckers75}) 
and early-type stars 
(\eg,~Vidal-Madjar et al.~\cite{avm88}) 
were unsuccessful (D is destroyed in stars).
Traub et al.~(\cite{traub74}) 
observed \ha\ in the Orion Nebula using three-etalon Fabry-Perot 
spectrometers and reported D/H upper limits. 

Here we report on spectra of Orion, secured at the Canada-France-Hawaii 
Telescope (CFHT). Emission lines detected in the blue wings of \ha\ 
and \hb\ are identified with \da\ and \db. A preliminary account 
was presented by H\'ebrard et al. (\cite{hebrard99}). 
Observations 
are described in 
Sect.~\ref{observations}, the identification and the origin of 
the lines in Sect.~\ref{identification} 
and \ref{origin} and the excitation mechanism in 
Sect.~\ref{fluo}.

\section{Observations, data reduction and results}
\label{observations}

Observations of the Orion Nebula (\object{M 42}, NGC~1976) were 
conducted at the 3.6m CFHT, using the Echelle spectrograph 
Gecko at the Coud\'e focus with a slit length of $\sim40$\arcsec.
The \ha\ and \hb\ spectral ranges were observed 
in October 1997 and September 1999 respectively. For \ha, 
the entrance slit was 1.2mm wide (3.5\arcsec\ on the sky), providing 
a resolution $R=\lambda/\Delta\lambda\simeq40\,000$ ($\sim7.5$~\kms); 
the detector was the $2048\times2048$ ``Loral~5'' thin CCD and 
the spectral range was 6544\AA\ - 6576\AA. 
For \hb, the slit was 0.8mm wide (2.3\arcsec) leading to 
$R\simeq50\,000$ ($\sim6$~\kms); 
the detector was the $2048\times4500$ ``EEV2'' thin CCD and 
the spectral range was 4832\AA\ - 4885\AA. 

The slit was centred 2.5\arcmin\ South of $\theta^1$~Ori~C 
(\object{HD 37022}), the brightest star of the Trapezium. 
The slit orientation was slowly rotating during the exposures 
(Coud\'e focus). 
Totals of 4.5 and 1.5 hours were devoted to \ha\ and \hb\ respectively,
divided in 30 -- 45~min sub-exposures. 
Small rotations of the grating were applied between \ha\ sub-exposures 
in order to disclose ghosts that may depend on grating setting. 
\ha\ was also observed at higher resolution ($R\simeq80\,000$) 
in the same area for 20~min. 
Finally, \hb\ was observed 
2.5\arcmin\ North and 20\arcsec\ South of $\theta^1$~Ori~C 
with shorter exposures and $R\simeq50\,000$.   
Bias, flats and Thorium-Neon lamp calibration 
exposures were secured regularly during the observations 
for each instrument configuration.

The spectra were reduced using MIDAS software. 
The steps of the data reduction were as follows:
(1)~bias subtraction; 
(2)~flat division; 
(3)~bad pixel and cosmic cleaning; 
(4)~summing the rows to transform the 2D-spectra into 
1D-spectra; 
(5)~wavelength calibration;
(6)~shift to the heliocentric frame; 
(7)~alignment of the different sub-exposures; and 
(8)~suming up of the sub-exposures. 
After shifting to the heliocentric frame, both \ha\ and \hb\ 
were fitted by a Gaussian on each sub-exposure. The standard 
deviation of the Gaussian peaks was less than 1~\kms. 
Sub-exposures were shifted to the average peak before summation 
in order to preserve the spectral resolution. 
An interfering signal, instrumental in origin, appeared in 
the \hb\ spectra, producing small oscillations in the dispersion 
direction, which slightly increased the noise level. 
Wavelengths are determined to better than 1.5~\kms\ and 1.0~\kms\ 
in the \ha\ and \hb\ final spectra respectively. 

\begin{figure}
\resizebox{\hsize}{!}{\includegraphics{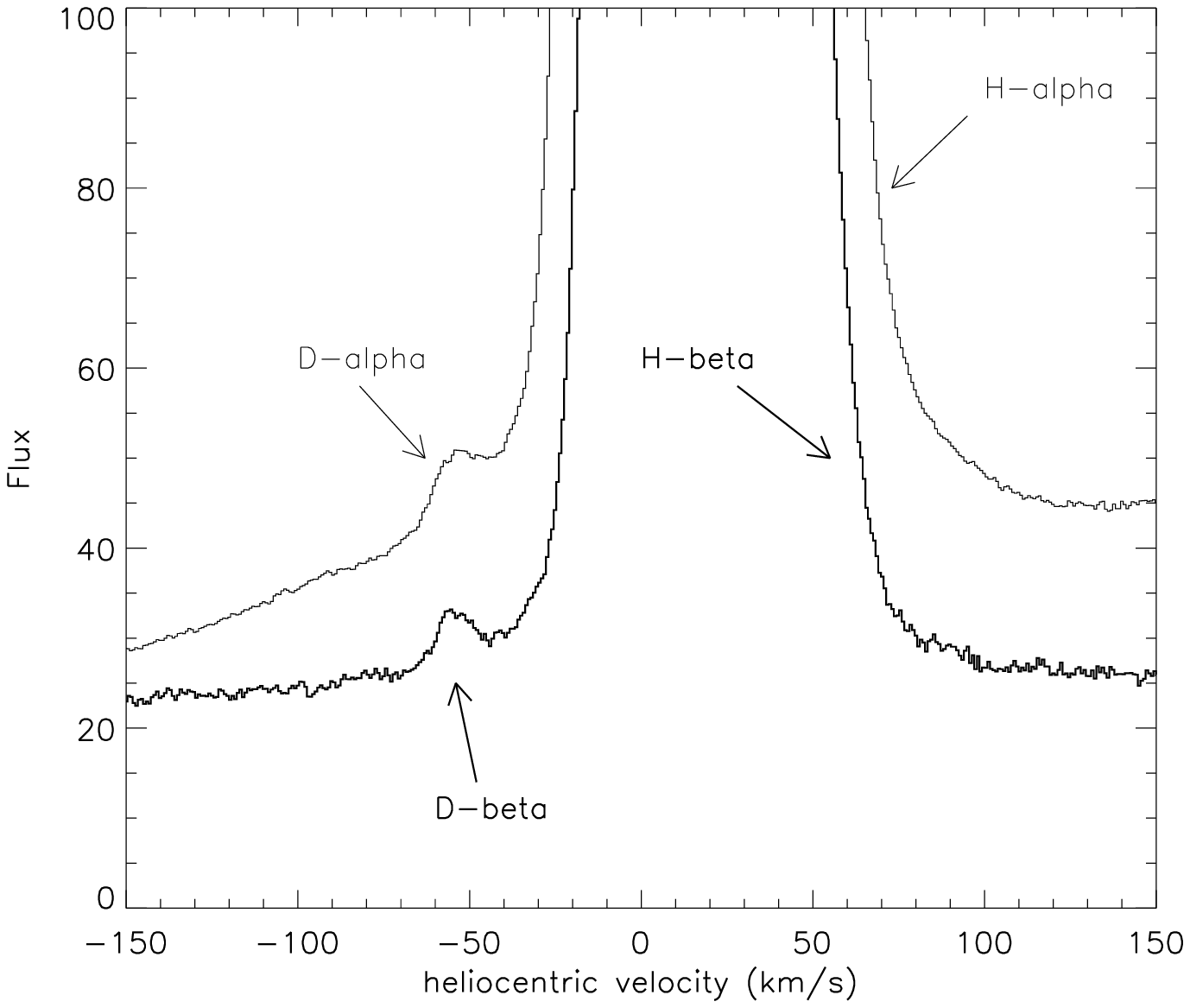}}
\caption[]{Vicinity of \ha\ and \hb\ in Orion, showing \da\ and \db\ 
in emission. 
The x-axis is in \kms\ relative to the rest wavelength of either \ha\ 
or \hb. The vertical scale corresponds to peak fluxes $7250$ and $2500$ 
for \ha\ and \hb\ respectively. 
Then $I$(\ha)$/I$(\hb)~=~2.91 and $I$(\da)$/I$(\db)~$\simeq1.10$. 
}
\label{fig_detection}
\end{figure}

Two weak emission lines are obvious 
in the blue wings of \ha\ and \hb\ (Fig~\ref{fig_detection}). 
Anticipating the conclusion of Sect.~\ref{identification}, the weak 
lines are already identified with \da\ and \db\ in Table~\ref{lines}, 
where results of Gaussian fits to \ha, \hb, \da\ and \db\ are given.
The full widths at half maximum (FWHM) of the deuterium lines are  
much smaller than those of the hydrogen lines. 
Relative fluxes (last row of Table~\ref{lines}) are based on the 
theoretical \ha/\hb\ ratio, thus implicitly correcting for reddening. 
For an H$^+$-weighted electron temperature 
(0.85$\pm$0.10)$\times$10$^4$K 
and electron density $\sim$ 5$\times$10$^3$~cm$^{-3}$
(\eg, Esteban et al.~\cite{esteban98}), the 
Case~B recombination ratio 
$I({\rm H}\alpha)$/$I({\rm H}\beta)$ is 2.91$\pm$0.03 
(Storey \& Hummer~\cite{stohum95}). 
Departure of this ratio from Case~B is expected to be much 
less than 1\% in this thick nebula for any reasonable dust content 
(Hummer \& Storey~\cite{humsto92}). 

\da\ and \db\ are seen all along the 40\arcsec\ slit. 
\db\ is present at all three positions. 
The velocity shifts between \db\ and \hb\ at 
2.5\arcmin~N, 20\arcsec~S and 2.5\arcmin~S of $\theta^1$~Ori~C 
are respectively 11.8, 9.1 and 10.0~\kms\ and the \db\ fluxes 
$4.2\pm1.1$, $2.3\pm0.6$ and 
$5.7\pm1.1$ (\hb$=10\,000$).

\begin{table*}
\caption[]{Gaussian fitted line profiles}
\label{lines}
\begin{tabular}{l|cc|cc}
\hline
Line identification & \ha & \da & \hb & \db \\
Rest wavelength (\AA) & $6562.796$ & $6561.010$ & $4861.325$ 
& $4860.003$ \\
\hline
Observed wavelength (\AA) & $6563.12\pm0.03$ & $6561.59\pm0.03$ 
& $4861.60\pm0.02$  & $4860.44\pm0.02$ \\
$v_{\odot}$ (\kms)$^a$ & $14.8\pm1.5$  & $26.5\pm1.5$ 
& $17.0\pm1.0$  & $27.0\pm1.5$  \\
FWHM (\kms)$^b$  & $32.0\pm0.5$  & $8.6\pm1.0$ 
& $32.1\pm0.5$ & $8.1\pm1.5$  \\
Relative flux$^c$ & 
$29\,100\pm300$ & $6.3\pm0.6$ & $10\,000\pm200$ & $5.7\pm1.1$ \\
\hline
\end{tabular}
\\
\\
$^a$ Heliocentric velocity. 

$^b$ Full width at half maximum corrected for instrumental width 
(original FWHM: 32.9, 11.4, 32.7, 10.1~\kms\ respectively).  

$^c$  Using the theoretical $I($\ha$)/I($\hb$)$ (see text), 
then $I($\da$)/I($\db$)=1.10\pm0.22$.

\end{table*}

\section{Identification of \da\ and \db}
\label{identification}

According to Table~\ref{lines}, the shift of both weak lines with 
respect to the hydrogen lines is $-71$~\kms\ whereas the isotopic 
shift of deuterium is $-81.6$~\kms. This significant difference 
forces us to consider alternative possibilities. 

$\bullet$
{\sl Spectral artifact?} \ 
No feature equivalent to these lines is present in either the 
red wings of \ha\ and \hb\ or the wings of \nii~$\lambda6548.05$\AA, 
observed simultaneously. Small rotations of the grating 
applied between sub-exposures resulted in no change in 
profile, position and intensity of the lines. Finally 
no such lines were detected in bright planetary 
nebulae we observed in September 1999 
(H\'ebrard et al.~\cite{hebrard00b}), using the same instrument. 
This also excludes possible sky-line emission. In fact, no sky-lines 
have been reported at these wavelengths. 
Thus, these lines are real features, specific to the Orion Nebula. 

$\bullet$
{\sl Unidentified process or element?} \ 
Attempts to find other identifications were unsuccessful. 
These lines cannot be scattered stellar emission 
[for example, \ha\ from $\theta^1$~Ori~C is variable 
(Stahl et al.~\cite{stahl96})], as lines from hot stars are broad. 
For the same reason, they cannot be Raman features. 
The wavelengths do not correspond to any known quasi-molecular line. 
The fact both lines have identical velocity shifts 
with respect to \hi\ practically restricts the possibilities to \hi\ 
and \di\ emission (no \ion{He}{ii} is detected in the Orion Nebula). 

$\bullet$
{\sl High-velocity hydrogen emitting structure?} \ 
Traub et al.~(\cite{traub74}) reported the detection of a line 
in the blue wing of \ha. This line may correspond to ours, although 
it was interpreted by these authors as high-velocity 
\hi\ emission, noting the existence of a similar component 
in [\ion{O}{iii}] (Dopita et al.~\cite{dopita73}). 
Indeed, our $R=80\,000$ spectrum shows a blue component 
in [\ion{N}{ii}] but with velocity shift only $\sim-22$~\kms. 
More importantly, {\it any component arising from the \hii\ region 
should have a minimum width corresponding to thermal broadening}. 
The thermal FWHM for hydrogen at 8500K 
is 20~\kms, much larger than 8~\kms\ (Table~\ref{lines}). 
The $-22$~\kms\ ionized hydrogen emission 
should be lost in the \ha\ and \hb\ wings. 

Nonetheless, the fluorescence mechanism proposed below 
to explain the \di\ line excitation (Sect.~\ref{fluo}) may 
a priori apply to \hi\ as well. 
One cannot formally exclude \ion{H}{i} fluorescence emission 
from a neutral, cold (thermal FWHM is $\sim7$~\kms\ at 10$^3$~K), 
high-velocity ($-74$~\kms\ LSR), low velocity dispersion ($<<10$~\kms) 
layer keeping the same kinematical properties over many arc minutes. 
However this is very unlikely as this hypothetical structure 
should in addition have a small column density 
(no other fluorescent line is seen at that velocity)  
{\sl and} lie sufficiently close to the Trapezium stars 
(fluorescence varies as the inverse square of the distance 
to the continuum source). 
The survival of a neutral thin shell against photoionization 
(no low-ionization material is interposed; see Sect.~\ref{origin}) 
is also in question. 
As a matter of fact, Cowie et al. (\cite{cowson79}) detected several 
components of high-velocity gas in absorption against $\iota$~Ori 
(a star located within half a degree of the region we observed), 
notably a component at $-68$~\kms, close to $-74$~\kms. 
According to these authors, this component 
corresponds to a very old highly ionized supernova remnant situated 
over 100~pc from the Trapezium, at any rate too far away to yield 
fluorescence. 

It can therefore be safely concluded that these lines are \da\ and \db. 
Kinematics (Sect.~\ref{origin}) brings out one more fundamental piece 
of evidence. 

\section{Origin of the lines}
\label{origin}

In the Orion Nebula, the \hii\ region is essentially matter bounded 
toward the observer and radiation bounded in the opposite direction
(\eg,~Rubin et al.~\cite{rubsim91}). 
The narrowness of the \ion{D}{i} lines implies that they must originate 
in a cold, localised region along the line of sight, that is behind 
the H$^+$ region, in the ``Photon Dominated Region'' (PDR) 
where deuterium is in atomic form. 

This is borne out by available information on velocities. The 
heliocentric velocity of the \ion{D}{i} lines is $\sim27$~\kms\ compared 
to $\sim16$~\kms\ for \ion{H}{i}. In a blue spectrum of the Trapezium region 
(Kaler et al.~\cite{kaler65}), the \ion{Si}{ii} lines 3856+63\AA, probably 
produced by fluorescence in the PDR, appear shifted by +11~\kms\ relative 
to the neighbouring \ion{H}{i} Balmer lines, a shift similar to the one 
found for \da\ and \db. Over a region close to the one we 
observed, Esteban \& Peimbert~(\cite{esteban99}) measured 
heliocentric velocities 
6.4$\pm$1.4, 14.0$\pm$2.0, 24.6$\pm$2.2 and 26.8$\pm$1.4~\kms\ for 
Ar$^{++}$, H$^+$, O$^0$ and N$^0$ respectively, 
tracing the free expansion of the \hii\ region, moving away 
from the molecular cloud. Over the same region, H\"anel~(\cite{hanel87}) 
found 20-23~\kms\ for [\ion{N}{ii}] [in agreement with our measurement 
$v($[\ion{N}{ii}]$)\simeq21$~\kms] and 23.5--28~\kms\ for [\ion{S}{ii}], 
thus encompassing the velocity we found for \da\ and \db. From 
millimeter and submillimeter observations, 
Hogerheijde et al.~(\cite{hoger95}) found velocities $\sim28$~\kms\ 
for different molecules. 
Observations thus clearly imply that the \ion{D}{i} lines could 
arise from the boundary of the \hii~region.

\section{Fluorescent excitation of \da\ and \db}
\label{fluo}

The narrowness of \da\ and \db\ allows the possibility to be 
excluded that the lines are excited 
by recombination in an ionized gas (Sect.~\ref{identification}). 
The H$^0$ column density of the PDR is 
$\sim$~10$^{22}$cm$^{-2}$ (Tielens \& Hollenbach \cite{tiehol85}), 
so the D$^0$ column density is $\sim$~10$^{17}$cm$^{-2}$ 
[assuming a typical \dshism$\simeq10^{-5}$] 
and the optical thickness in, \eg, \lybd\ is over 100. 
Since the \di\ emission is confined to a layer coincident in velocity 
with that of the PDR (Sect.~\ref{origin}) and since the dust opacity 
there is orders of magnitude less than the \lybd\ opacity, 
fluorescence from the Lyman lines is a viable process to produce 
the deuterium Balmer lines. 

The UV continuum is dominated by $\theta^1$~Ori~C, whose effective 
temperature is close to $4\times10^4$K (Rubin et al.~\cite{rubsim91}). 
Let us assume that both the ionization of the 
\hii\ region and the deuterium fluorescence are due to a 
$4\times10^4$K black body and that half the ionizing photons 
escape from the \hii\ region in the matter bounded directions 
(Rubin et al.~\cite{rubsim91}). 
If each photon impinging on the PDR at the \lybd\ wavelengths 
ultimately produces a \da\ photon by scattering on D$^0$, and 
only \lybd\ photons lead to \da\ excitation 
(neglecting cascades), then the flux ratio 
$I$(\da)/$I$(\ha) is about $1.5\times10^{-4}\times(\Delta v/5$\kms), 
where $\Delta v$ is the full velocity width of the zone where \da\ is 
effectively excited. According to Tielens and Hollenbach (\cite{tiehol85}), 
the turbulent pressure in the PDR corresponds to 
$\Delta v\simeq5$~\kms\ and according to Table~1, $\Delta v$ is 
probably less than 8~\kms. The rather good agreement of this very 
coarse estimate with the observed value $2.2\times10^{-4}$ (Table~1) 
is fortuitous. This estimate 
may be wrong in different ways. Many Ly$_{\rm D}$ lines can a priori 
absorb primary photons and feed \da\ by cascades, then leading to 
an overestimation. Conversely, part of the Ly$_{\rm D}$ photons are 
absorbed by dust and/or reflected back to the \hii\ region. 
Also, our estimate is global in character and 
our particular line of sight may not intercept identical fractions 
of the \hii\ region and the PDR. Most importantly, 
the stellar continuum may be depleted in the vicinity of the 
Ly$_{\rm H}$ lines, particularly for the first members of the series. 
Nonetheless, since these different effects tend to partially 
compensate one another, the above agreement indicates that the 
assumption of UV continuum fluorescence leads 
to the correct order of magnitude for the \da\ flux. 

Unlike \da/\ha, the flux ratio \da/\db\ is little sensitive to 
aspect effects. Assuming a ratio of visual extinction to column density 
of hydrogen nuclei $A_V$/$N_{\rm H} = 5\times$10$^{-22}$~mag$\,$\cm2\ 
and $A_{FUV}$/$A_V = 5$ (Tielens and Hollenbach \cite{tiehol85}), 
the ratio of \lybd\ opacity to dust opacity is $\sim$~240. 
Only for large principal quantum numbers should dust absorption 
decrease \di\ fluorescence (the photoexcitation cross section goes 
roughly as $n^{-3}$). Since the stellar continuum should be about flat 
over the Lyman line range (except possibly in the vicinity of strong 
lines), some insight into the excitation process 
can be gained by assuming that all Ly$_{\rm D}$($n$) lines, 
with principal quantum number $n$ up to some given $n_0$, convert 
identical numbers of UV photons by fluorescence and that 
no fluorescence occurs for $n > n_0$. Then, working out the 
cascades, the theoretical \da/\db\ flux ratio is about 1.28, 1.41 
and 1.51 for $n_0 =$ 4, 5 and 6 respectively and tends to level off 
for larger $n_0$'s. Only for $n_0=4$ is this simple 
description compatible with the observed value $1.10\pm0.22$ (Table~1). 
Since one would expect that a relatively large number of Ly$_{\rm D}$ 
lines should contribute to the excitation, the suggestion is that 
one of the above assumptions was oversimplified or some significant 
process has been overlooked. For example, the stellar continuum 
is probably depleted in the wings of the Ly$_{\rm H}$ lines and 
the vicinity of \lyb\ is likely to be most affected, 
then selectively reducing the \da\ emission. 
Observing higher deuterium Balmer lines is essential before attempting 
any detailed modeling. 

Note that the \da/\db\ of Table~\ref{lines} 
was obtained assuming that the reddening correction was the same 
for the \hi\ and \di\ emitting zones. If extinction internal to the 
nebula is significant, then the actual (de-reddened) \da/\db\ ratio 
will be even smaller since the PDR, where the \di\ lines come from, 
is more deeply embedded than the \hii\ region. 

Dust absorption will dominate Ly$_{\rm D}$ fluorescence for sufficiently 
large $n$, the deuterium Balmer decrement then changing from very flat 
to very steep. This break can lead to a D/H value inasmuch as the 
dust opacity per hydrogen nucleus is known. On the other hand, 
fluorescence lines from species co-extensive with D$^0$ 
including \ion{O}{i} and \ion{Si}{ii} can provide 
independent information on the primary continuum flux 
and on the competition of line scattering 
with dust absorption for photons. 
\ion{O}{i} fluorescence lines have been detected long ago in Orion 
and the excitation process was established by Grandi~(\cite{grandi75}). 
Comparing \di\ and \oi\ lines may lead to a D/O abundance ratio. 
Detailed observations and proper modeling of many deuterium Balmer 
lines and other fluorescence lines appear as a potentially 
accurate means to determine D/H in \hii\ regions. 

\section{Conclusions}

Deuterium is identified for the first time in a nebula from optical 
spectroscopy. The excitation mechanism of the observed lines, 
\da\ and \db, is continuum fluorescence from Ly$_{\rm D}$ lines in 
the PDR. Considering the saturation of the first Ly$_{\rm D}$ lines 
and the possible influence of the neighbouring Ly$_{\rm H}$ lines, 
observing the full deuterium Balmer series is essential 
to obtain a D/H value from optical data and appears feasible, 
at least in Orion. 

An optical determination of D/H in \hii\ regions would allow 
to check existing \dshism\ values and obtain D/H in 
low-metallicity extragalactic \hii\ regions, where the 
deuterium abundance should be close to its primordial~value. 

The large photoexcitation cross section of the first Lyman 
lines makes deuterium Balmer fluorescence a sensitive way 
to {\sl detect} deuterium in nebulae, leading for example 
to stringent upper limits to D/H in planetary nebulae
(H\'ebrard et al.~\cite{hebrard00b}), where deuterium is 
depleted.

\end{document}